\documentclass[12pt,preprint]{aastex}
\usepackage{emulateapj5}

\newenvironment{inlinefigure}{%
\def\@captype{figure}%
\noindent\begin{minipage}{0.999\linewidth}\begin{center}}
{\end{center}\end{minipage}\smallskip}

\shorttitle{Steward Sucks!}
\shortauthors{Barnes \& O'Brien}

\begin{document}

\title{Superiority of the Lunar and Planetary Laboratory (LPL) over
Steward Observatory (SO) at the University of Arizona}

\author{Jason W. Barnes, D. P. O'Brien, J. J. Fortney, Terry A. Hurford}
\affil{Department of Planetary Sciences LPL, University of Arizona, Tucson, AZ, 85721}
\email{jbarnes020@c3po.lpl.arizona.edu}

\begin{abstract} 

LPL dominance over Steward is demonstrated through several lines of evidence
including observations and modelling.  The decrease in Steward coolness is
attributed to the departure of interesting graduate students from Steward that
enrolled at LPL.

\end{abstract}

\keywords{interdepartmental rivalries --- pranks:April Fools --- 
 departments:individual(Steward) --- departments:individual(LPL)}

\section{INTRODUCTION}

Both the astronomy department and the planetary sciences department at the
University of Arizona, known as Steward Observatory (SO) and the Lunar and
Planetary Laboratory (LPL) respectively, have enjoyed scientific success over the
past $50$ years.  In more recent times however, SO has experienced a serious
decline in the quality of its graduate students and their education, thus leading
to the inevitable conclusion that LPL is the far greater institution for the
production of scientifically useful dotoral candidates.

\section{EVIDENCE}

A competitive volleyball tournament was held during 2001 between graduate
participants from both LPL and SO.  The match terminated in a rout of epic
proportion.  Steward graduate students were so humiliated so as not to ever
return to the game of volleyball.  We quit 'cuz we got bored.

\begin{inlinefigure}
\bigskip
\centerline{\includegraphics[width=1.0\linewidth]{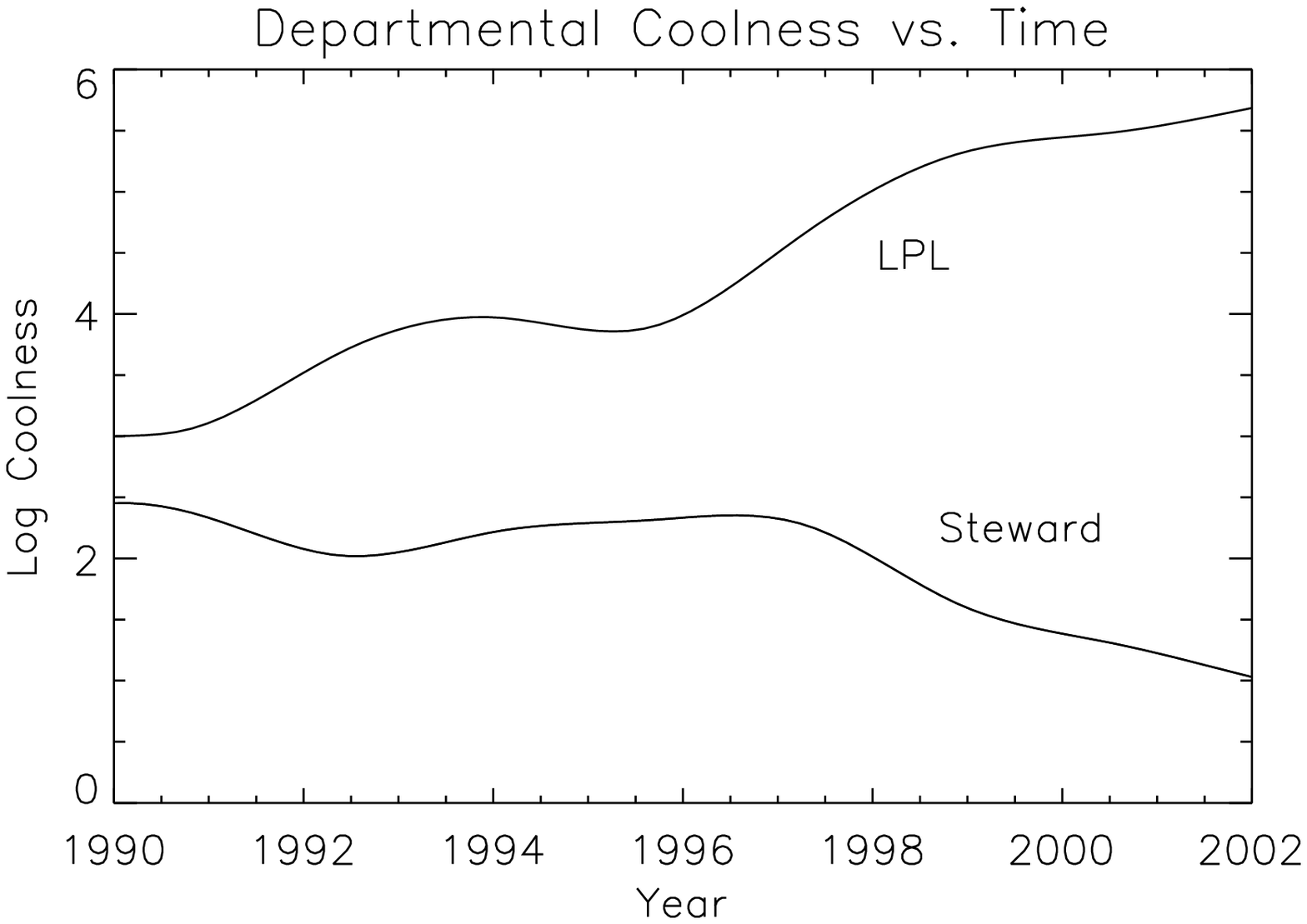}}
\bigskip
\end{inlinefigure}

\vspace{-0.3in}

{\tiny\emph Departmental coolness as a function of time over the last decade for
both LPL and Steward Observatory.  The y-axis is measured in coolombs. 
Particularly note the divergence in coolness values between the two institutions
in 1998 -- this is believed to be the result of the defection of Terry Hurford
from SO to LPL at that time.}

In addition, the above facts are buttressed by the extreme and total humiliation
that Astronomy Department graduate students suffered at the hands of the Planetary
Science grads on April 1, 2001 (Bieging, private communication).  On that date the
Steward Observatory email list was informed of the impending annexation of their
department by the Lunar and Planetary Laboratory due to budget constraints, and
signs attesting to this new status were emplaced around the building.  So
infinitessimal are/were the intellectual and technical capabilities of SO graduate
students that not only did one of these signs remain in position for over 1 week,
but that in fact no counter-pranks have been executed for the past 1 year.  We
argue that these events imply the complete subordination of SO grads to the will
of the LPL graduate population.

The evidence presented in this section is summarized in Table 1.

\vspace{0.3in}
\begin{tabular}{l|l}
\hline
10 & Steward grads live in Hawthorne House.\\
 9 & Who really cares about high-Z quasars anyway? \\
 8 & Revenge of the Nerds -- 'nuff said.\\
 7 & Squirrel-loving, peyote-smoking hippie \\
   & freaks don't try to break our mirrors.\\
 6 & Bratfest.\\
 5 & Lack of pimply undergraduate majors.\\
 4 & Able to waste \$250,000,000 in a single\\
   & botched landing attempt.\\
 3 & LPL professors unafraid to take spectra\\
   & of own urine.\\
 2 & Steward graduate students not nearly\\
   & creative enough to have thought of this.\\
 1 & Field trips.\\
\hline

\end{tabular}

\acknowledgements
{\scriptsize
The authors wish to acknowledge an anonymous reviewer's extraordinarily harsh
review -- no doubt the rest of our careers will seem easy in comparison to what
this person made the publication of this paper like for us.  DO is funded through
the illigitimate sale of `crack' cocaine to Steward Observatory graduate
students.  JB isn't really funded at all, he's just leaching off the rest of the
department.}

\end{document}